\input{aipcheck}

\documentclass[
    ,final            
  ]
  {aipproc}

\layoutstyle{6x9}

\begin{document}

\title{Scalar field and QCD constraints in Nuclear Physics}


\author{M. Ericson}{
  address={Universit\'e de Lyon, Univ.  Lyon 1, 
 CNRS/IN2P3,\\ IPN Lyon, F-69622 Villeurbanne Cedex\\
and Theory division, CERN, CH-12111 Geneva }
}

\author{G. Chanfray}{
  address={Universit\'e de Lyon, Univ.  Lyon 1, 
 CNRS/IN2P3,\\ IPN Lyon, F-69622 Villeurbanne Cedex}
}

\begin{abstract}
Relativistic theories of nuclear matter are discussed in a new pespective. First the chiral character
of the scalar nuclear field is introduced in the framework of the linear sigma model. With the assumption
that the nucleon mass originates in part from the coupling to the quark condensate it is possible to relate
the optical potential for the propagation of the scalar field to the QCD scalar susceptibility of the nucleon, 
on which indications exist from the lattice  evolution of the nucleon mass with the quark mass. Constraining
the parameters of the nuclear scalar potential by the lattice expansion parameters
a successful description of the nuclear saturation properties can be reached.  
\end{abstract}

\maketitle


In this meeting devoted to the scalar mesons it is appropriate to talk of nuclear physics and I am one speaker in charge of this task.  Indeed it is clear that the existence of a scalar meson coupled to nucleons has big implications for nuclear physics. This idea has been exploited in the relativistic models of the nuclear binding of Walecka and Serot \cite{SW86}. Here the binding results from the balance between the attractive sigma exchange and the repulsive omega exchange. This model had impressive successes, in particular for what concerns the spin-orbit coupling. The main new idea which has emerged since the introduction of this model is the concept of a nucleonic response to the scalar field proposed by Guichon \cite{G88,GSRT96}. The nucleon is not inert but it responds to the presence of the field. This concept in itself is not new and occurs for instance in pion propagation. The original aspect of the quark meson coupling model (QMC) is that it implies the quark struture of the nucleon. We will come back to this model. Here we propose the existence of a link between the scalar parameters of the scalar potential and the QCD lattice data \cite{EC07} and I will try to convince you that it is natural and to be expected.

If one talks about a link between nuclear physics and QCD, it is natural to invoke effective theories which simulate QCD at low energies in terms of hadronic degrees of freedom and we  use as a starting point the linear sigma model. Here chiral symmetry is implemented with the introduction of two fields, the $\sigma$, scalar isoscalar,  and the $\pi$, pseudoscalar isovector, which are chiral partners. In this model the part of the lagrangian
which breaks chiral symmetry, which in QCD is $ {\cal L}_{sym break}^{QCD}= -2\,m_q \,{\bar q q}$, is replaced by 
$ {\cal L}_{sym break}^{model}=f_{\pi}\,m_{\pi}^2\,\sigma$.
Due to the spontaneous symmetry breaking, the expectation values of the order parameter are non vanisishing,
with the vacuum values $\langle\bar q q\rangle_{vac}$ in QCD and $\langle\sigma\rangle= f_{\pi}$.
The equivalence of the symmetry breaking parts of the QCD and model lagrangians
implies the following equivalence between the quark density and the scalar field~: 
$${\bar qq(x)\over{\langle\bar qq\rangle_{vac}}}={\sigma(x)\over f_{\pi}}.$$ 
Taking expectation values we find that the quark condensate in the nuclear medium is governed by the expectation value of the sigma field. 
 
We have now to incorporate the concept of the nuclear scalar field in this theory. A natural and simple 
identification would be  that of the nuclear mean scalar field with the expectation value 
$\delta\langle \sigma\rangle=\langle \sigma\rangle\,-\,f_\pi$.  It would make life simple because a unique quantity
$\delta\langle \sigma\rangle $  would govern the evolutions of the condensate and of the effective nucleon mass in the nuclear medium, $M_N^*$, with~:
\begin{equation}
{M_N^*\over M_N} ={\langle\bar qq(\rho)\rangle \over {\langle\bar qq\rangle_{vac}}} = 1+ {\delta\langle \sigma\rangle\over f_{\pi}}
\simeq 1-{\Sigma_N \rho_N^S\over {f_{\pi}^2 m_{\pi}^2}}
\end{equation}
where $\rho_N^S$ is the nucleon scalar density and $\Sigma_N$ the nucleon sigma commutator.
As the condensate evolution is known for  independant nucleons from the value of $\Sigma_N$, as written  above, the mean scalar field would be known to first order in density, in a model independent way, with a value of $\simeq 30\, MeV$ at
normal density. Unfortunately this simple identification is not legitimate, as emphasized by Birse \cite{B96}. The pionic contribution to the nucleon sigma commutator contains non analytical terms in $m_{\pi}^3$. It follows that the NN interaction which governs the mass evolution would have terms of order $m_{\pi}$, which is not allowed by the chiral constraints.  Of course it is always possible to add other exchanges, involving in particular two-pion states, in such a way that the unwanted terms cancel. But this is a very cumbersome procedure which is not practiced. We have proposed \cite{CEG01} a short cut with the suggestion that it is interesting to go from the cartesian coordinates, 
$\sigma$ and $\pi$, to polar coordinates,  $S$ and $\phi$ according to~:
$\sigma\, + \,i\vec\tau\cdot\vec\pi=S\,U=(f_\pi\,+\,s)\,exp\left({i\vec\tau\cdot\vec\phi/ f_\pi}\right)$,
where $S$ is associated with the radius of the chiral circle and
$\phi/f_{\pi}$ the polar angle. By doing so we go from a linear representation to a non linear one. Contrary to the usual procedure of the non linear model we do not freeze the radius of the chiral circle but we allow for a change in the nuclear medium, where it goes from its vacuum value, $f_{\pi}$, to $f_{\pi} \,+\,s$. We have suggested to link the nuclear field to this radial mode with the identification of the background mean nuclear scalar field with the mean value $\bar s$. Since we are in a non-linear representation all chiral constraints are automatically satisfied. 
We point out  that the bare masses of the $\sigma$ and the $s$ are identical but these two fields differ by their  off-shell couplings to two-pion states. In particular, due its derivative couplings, the $s$ field decouples  from low energy pions. 
 
Note that the extreme simplicity reached in the previous identification is lost. We have now two scalars, one
 $\delta\langle \sigma\rangle $ which is not invariant under chiral transformations governs the condensate evolution, and $\bar s$ which is a chiral invariant, governs the nucleon mass one. However they are not unrelated. Expressing the sigma field in terms of the polar coordinates we obtain for the condensate evolution, keeping only the lowest order terms~:
\begin{equation}
\label{Eq1}
{\langle \overline{q}q\left( \rho \right)\rangle\over {\langle\bar qq\rangle_{vac}}}=1\,+\,{\bar s\over f_{\pi}}\,-\,\frac{\left\langle \phi ^{2}\right\rangle }{2\,f_{\pi }^{2}}.
\end{equation}
Here the quantity  $\left\langle \phi ^{2}\right\rangle $ represents the scalar density of virtual pions in the nuclear
cloud divided by the pion mass. For the nucleon mass instead~:
\begin{equation}
\label{MASS}
{M^*_N(\rho) \over M_N}=1\,+\,{\bar s\over f_{\pi}}.
\end{equation}
The nuclear scalar mean field still influences the condensate. However in order to extract it from the condensate, the effect of the pion cloud has to be separated out, which introduces an unavoidable model dependence.

So we are now at peace with chiral symmetry but we face another problem. This chiral invariant nuclear scalar field is subject to the chiral dynamics. The mexican hat potential of the chiral theory contains a three scalar coupling term which produces a lowering of the sigma mass in the medium~:
\begin{equation}
m^{*2}_{\sigma}  
= m^{2}_{\sigma}\,-\,{3\,g_S \over f_\pi}\,\rho_N^S,
\label{MSIGMA}
\end{equation}
or equivalently for the existence of a tadpole $\sigma N$  amplitude~:
\begin{equation}
T_{\sigma N}= -3\,g_S / f_{\pi }\label{TSN}.
\end{equation}
Here $g_S$ is the $\sigma$-nucleon coupling constant.
This lowering is a large effect, about $30\%$  decrease of the mass at normal density. It means more attraction with increasing density, such that a collapse occurs, instead of saturation, as pointed out long ago by Kerman and Miller \cite{KM74}. The $\sigma$ 
model simply does not provide a viable theory of nuclear matter. Phenomenologically, the problem can be cured with the introduction of a response of the nucleon to the scalar field, in the form of a polarizability coefficient, $\kappa_N$
such  that the nucleon mass evolves, as in QMC, according to~: 
\begin {equation}
M_N^* = M_N \,+\, g_S\, \bar s \,+\,{1\over 2}\,\kappa_{N}\,\bar s^2.
\label {MSTAR}
\end {equation}
With a positive value of the response, repulsion is introduced and it is possible to reach saturation. For us \cite{CE05}, with an empirical value of $\kappa_{N}$ such that approximately two thirds of the tadpole effect are cancelled,  saturation is possible.  But phenomenology is not our aim here and no link with QCD emerges. The link will appear with the study of the QCD scalar polarizability, $\chi_S$, defined as 
$\chi_S =  \partial \langle\bar q q\rangle / \partial m_q$,
the derivative of the order parameter $\langle\bar q q\rangle$ with respect to the bare mass parameter $m_q$ which breaks explicitely chiral symmetry. 
For the nuclear medium we define the susceptibility $\chi_S^A$ in such a way that the vacuum value is subtracted off, and only the condensate evolution enters. The susceptibility represents also the propagator of the fluctuations of the order parameter, $\delta \langle\bar q q\rangle(x)$, which in the sigma model is simulated by  $\delta \sigma(x)$. The susceptibility is thus expressed in the model in terms of the $\sigma$ propagator, with~:
\begin{equation}
\chi _S^A\,=\,{2\,{\langle\bar q q\rangle_{vac}^2\over f_{\pi }^2} }
\left[{1\,\over  m^{*2}_{\sigma }}
\,-\, {1\over  m^{2}_{\sigma}  }\right] .
\label{CHIAS}
\end{equation}\
where $m^*_{\sigma}$ is the in-medium sigma mass, as modified by the tadpole term. Expanding the r.h.s. of 
eq. \ref{CHIAS} to first order in the nucleon density we obtain~:
\begin{equation}
\chi _S^A\,=
\,2\,{\langle\bar q q\rangle_{vac}^2\over f_{\pi }^2\, m_{\sigma}^4}\,\left({-3\,g_S\over f_{\pi}}\right)\, \rho_S^N +..... 
\label{EXPANSION}
\end{equation}
The meaning of this linear term in density is clear.  We are dealing with the modifications brought  into the vacuum susceptibility by the presence of the nucleons. Among these, part arises from the susceptibility of the nucleons which individually respond to a modification of the quark mass. The
factor in front of the density therefore represents the scalar susceptibility of a free nucleon which is found as \cite{CEG03}~:
\begin{equation}
(\chi^N_S) \,=\,2\,{\langle\bar q q\rangle_{vac}^2\over f_{\pi}^2\, m_{\sigma}^4}\,\left({-3\,g_S\over f_{\pi}}\right). 
\label {CHISN}
 \end{equation}
The linear sigma model thus predicts the existence of a negative component in $\chi_S^N$ which is associated with the scalar field and linked to the tadpole amplitude. After having obtained this result it was possible to derive it directly  in the model, as the derivative of the nucleon scalar charge with respect to the symmetry breaking parameter but in this case the connection to the tadpole amplitude does not emerge.
 
This is the sigma model prediction. In real life, is there any evidence for this component~?
Possibly, in the lattice simulations of the evolution of the nucleon mass with the quark mass (equivalently the squared pion mass)~
\cite{TLY04,TGLY04}. These results do not cover the physical mass but only a region of the quark mass above
 $\simeq 50\,MeV$ with several data points above this value. In order to extract the physical nucleon mass, an extrapolation has to be performed. For us the fact of having several data points at varying quark masses is interesting because the successive derivatives of the nucleon mass with respect to the quark mass provide the nucleon sigma commutator and the scalar susceptibility, according to :
\begin{equation}
\Sigma_N = m^2_{\pi} \,{\partial M\over \partial m^2_{\pi }},
\label{SIGMATOTALLATTICE}
\end{equation}
and~:
\begin{equation}
\chi_S^ N~= 2\,{\langle\bar q q\rangle_{vac}^2 \over 
f^4_{\pi} }
{\partial \over \partial  m^2_{\pi }}\left({\Sigma_N \over 
m^2_{\pi }}\right)
  \label{CHILATTICETOTAL}
\end{equation}
These however are total values which include the pionic contributions, which should be removed for
our nuclear physics purpose. Fortunately the pion loop contribution to the nucleon mass evolution
has been separated out with great care in the works of ref. \cite{TLY04,TGLY04}. The reason is that it contains non analytical terms in
the quark mass which prevent a small mass expansion.  
The pion contribution depends on the $\pi N$ form factor. Different form factors have been used with an adjustable cutoff parameter $\Lambda$, which is fitted. The non-pionic mass is expanded in $m_q$ (equivalently in $m_{\pi}^2$) as follows~:
\begin {equation}
M_N(m^{2}_{\pi}) = 
a_{0}\,+\,a_{2}\,m^{2}_{\pi}\, 
+a_{4}\,m^{4}_{\pi}\,+\,\Sigma_{\pi}(m_{\pi}, \Lambda).
\label{EXPANSION}
\end{equation}
The best fit values of the parameters $a_{4}$ and $a_2$ show little sensitivity to the shape of the form  factor, with 
$a_4 \simeq-\,0.5\, GeV^{-3}$, while $a_2 \simeq 1.5\,GeV^{-1}$ \cite{TGLY04}.
 From the expansion of eq. (\ref{EXPANSION}) we deduce ~:
\begin{equation}
\Sigma_N^{non-pion} = 2\,m_q \,Q_S = m^2_{\pi} \,{\partial M\over \partial m^2_{\pi }}
= a_2\, m^2_{\pi}\, +\, 2\,a_4\,  m^4_{\pi}~\simeq a_2\, m^2_{\pi}= 29\,MeV \,.
\label{SIGMALOT}
\end{equation}
It is largely dominated by the $a_2$ term. The quantity $Q_S$ defined above is the total scalar quark number of the nucleon, that we denote also as the nucleon scalar charge. In turn the nucleon susceptibility is~:
\begin{equation}
\chi_S^{ N, non- pion}~= 2\,{\langle\bar q q\rangle_{vac}^2 \over f^4_{\pi} }
{\partial \over \partial  m^2_{\pi }}\left({\Sigma_N^{non-pion} \over m^2_{\pi }}\right)=  
{\langle\bar q q\rangle_{vac}^2 \over f^4_{\pi} }\,4\,a_{4}
 ~\simeq -5.4\, GeV^{-1} 
\label{CHILATTICE}
\end{equation}
The non-pionic susceptibility is found with a negative sign, as expected from the sigma model. If signs are right, are the magnitudes also compatible with the sigma model~? In this case the scalar charge is~: 
$$Q_S={ \langle\bar q q\rangle_{vac}\, g_S \over f_{\pi}\,m_{\sigma}^2}$$
As for the suceptibility the identification of the sigma model result and the lattice one leads to:
\begin{equation}
 (\chi_S^N)^{non- pion}={2 \,(Q_S)^2\over g_S^2}\,\left({-3\,g_S\over f_{\pi}} \right)
 =~{\langle\bar q q\rangle_{vac}^2 \over f^4_{\pi} }\,4\,a_{4} 
\end{equation}
which, using the relation (\ref{SIGMALOT}) between $Q_S$ and $a_2$ gives~:
\begin{equation}
-a_4={3\over2} \,{(\Sigma_N^{non-pion})^2 \over g_S\,f_\pi\,{m_\pi}^4} 
\simeq {3\over 2}\,{a^2_2 \over g_S \,f_{\pi}} = 3.5\, GeV^{-3},
\end{equation}
seven times larger than the lattice value, $-a_4= 0.5 \, {\rm GeV}^{-3}.$
The sigma model is contradicted also by the value of the expansion parameter $a_4$, a new failure for this model~? No, it is in fact the same as the previous one since the susceptibility and the tadpole amplitude are proportional and the sigma model prediction fails for both, as it should. This full consistency between the nuclear physics and the QCD lattice results is even gratifying. The question is if we can go beyond this failure. There is a need for a compensating terms. The common cure is found in confinement, which introduces a positive response of the nucleon to the scalar field, as in QMC. This model is a pure bag model in which the nucleon mass entirely arises from confinement. Moreover the chiral caracter of the nuclear scalar field is not discussed. Here instead we will assume that the nucleon mass originates in part from confinement and in part from the condensate. In addition we keep the assumption that the nuclear scalar field is the chiral invariant field discussed previously which influences the quark condensate.

The first question is~: in this mixed situation what is the relation between the nucleon scalar susceptibility and the $\sigma N$ scattering amplitude~? We will illustrate this point in a hybrid model of the nucleon, similar to the one introduced by Shen and Toki \cite{ST99}  
which  consists in the following.  Three constituant quarks,  described by the Nambu-Jona-Lasinio model, move  
in a non-perturbative vacuum, their mass M is generated from the coupling to the quark condensate. They are kept together by a central harmonic potential which mimics confinement and the effect of the color string tension. Although oversimplified the model  provides an intuitive picture  of the role played by confinement. The nucleon mass, because of the confining force, becomes $3\,E(M)$, the 
M dependence being fixed by the type of force. For illustration we take a harmonic potential of the form 
$(K/4)(1+\gamma_0)\,r^2$, which leads to the expression~:
\begin{equation}
M_N =3\,E=3\left( M+{3\over2}\sqrt{K\over E+M} \right) \,.
\end{equation}
It is increased as compared to the value, $3\,M$, for three independent quarks. With our assumptions the presence of the mean nuclear scalar field in the medium modifies the condensate and hence affects the mass $M$. The derivative  $\partial M /\partial{\bar s}= g_q = M/f_{\pi}$
has the non-vanishing value of the NJL model. The scalar coupling constant of the nucleon is~:
\begin{equation}
 g_S ={\partial M_N\over \partial {\bar s}} = 3\, g_q \, {\partial E\over \partial M}.
\end{equation}
The nucleon scalar charge, $Q_S$, writes~:  
\begin{equation}
Q_S ={3\over 2}\, {\partial E\over \partial m_q}  
={3\over 2} \,{\partial E\over \partial M} \, {\partial M \over \partial m_q}
\end{equation}
where $3\,\partial E/ \partial M=3\,(c_S=3\,E+3M/ 3E+M)$ is the scalar number of constituant quarks. As $E>M$, $c_S<1$, it is reduced as compared to a collection of three independent  quarks. The nucleon scalar susceptibility, $\chi^N_{S}$, given by the next 
derivative, is composed of two terms arising respectively from the derivative of $c_S$ and from that of  $\partial M/ \partial 
m_q$~:
\begin{equation}
\chi^N_{S} = {\partial Q_S\over \partial m_q}= {3\over 2} 
\left[ {\partial c_S\over \partial M}\left({\partial M\over \partial m_q}\right)^2
+ c_S \, {\partial^2  M\over \partial^2  m_q^2} \right] 
\quad\hbox{with}\quad
 {\partial c_S\over \partial M}={24 \,(E^2- M^2)\over {(3\,E+M)}^3}\,.\label{DOUBLE}
\end{equation}
Notice that this last derivative vanishes in the absence of confining force, when $E=M$, and that it is positive since $E>M$.  
Therefore the first part of the expression of $\chi^N_{S}$ represents the part of the susceptibility originating in confinement which is positive as in QMC. While instead the other term, proportional to the susceptibility of a constituant quark,
$\partial^2  M / \partial^2  m_q^2$, is negative. Thus a compensating effect is possible. 

Is there a similar compensation in the $\sigma N$ scattering amplitude~? Two terms as well contribute to $T_{\sigma  N}$. One is the tadpole term on the constituant quarks. For each constituant quark the tadpole amplitude is $t_{\sigma N} = -3\,g_q/f_{\pi}$.  Multiplying by the scalar number of constituant quarks which is $3\,c_S$, the tadpole amplitude for the nucleon writes  ${T_{\sigma N}}^{tadpole} = -3\,g_S/f_{\pi}$, 
the same expression as in the linear sigma model. We can compare this amplitude to the part of the nucleon 
susceptibility arising from the constituant quark susceptibility, second term on the r.h.s. of the expression (eq. \ref{DOUBLE}) of $\chi^{N}_{S}$. The constituant quarks as described by the NJL model obey at the quark level, the same relations as in the
linear sigma mode. In particular the relation between the quark susceptibility and the tadpole amplitude is :
$${\partial^2  M\over \partial m_q^2}= 2\left({\partial  M\over \partial  m_q}\right)^2\,\left({-3\, g_q\over f_\pi}\right).$$
Multiplying both members by $3\,c_S$ we find that the nucleonic suceptibility originating from that of the constituant quarks one is related to the tadpole amplitude by the same ratio $2\,Q_S^2/g_S^2$ as  previously. As for the part originating in confinement, the amplitude $\kappa_N$ is obtained as the second derivative of the nucleon mass with respect to the scalar field~: 
\begin{equation} 
\kappa _{N} = 3\,
{\partial^2 E\over \partial \bar s^2}= 3\, {\partial  c_S\over \partial M}\left({\partial M\over \partial \bar s}\right)^2\,.
\end{equation}
The ratio, $r_m$,  between the part of the nucleon scalar susceptibility  due to confinement and $\kappa_{N}$ is
\begin{equation}
r_m={1\over 2}\,{({\partial M\over \partial m_q})^2\over ({\partial M\over 
\partial s})^2}
= {2 \,Q_S^2 \over g_S^2},
\end{equation}
the same ratio $r$ as was previously found. Adding then the two components we obtain  : 
\begin{equation} 
\chi _{S}^N =  {2 \,Q_S^2 \over g_S^2}\,T_{\sigma N}^{total} 
\label{TRUTH}
\end{equation}
The introduction of confinement has preserved  the relation between the scalar susceptibility and the $\sigma N$ amplitude.
Numerically our model is not successful. It does not produce enough cancellation of the tadpole effect but it is conceptually important to understand the role of confinement. The relation between the susceptibility and the $\sigma N$ amplitude established in the framework of this model is actually independent of the particular form of the function $E(M)$. A more sophisticated model of the nucleon is needed for a quantitative description.  

The  relation (\ref{TRUTH}) allows to go to the next step of our approach which consists in using the lattice data to fix or at least constrain the nuclear physics parameters of the scalar interaction part.  In this approach for instance the medium effects in the propagation of the scalar field directly follow from the eq.(\ref{TRUTH}) and we can write the $\sigma$ propagator as~: 
\begin{equation}
-(D^*_{\sigma})^{-1}=m_{\sigma}^2\,+\,g_S^2\,{2\,a_4\over a_2^2}\,\rho_S^N = =m_{\sigma}^2\, -\,0.5\, {g_S\over f_{\pi}}\,\rho_S^N.
\end{equation}
The sigma mass is considerably stabilized with respect to the pure sigma model, where the coefficient in front of the last factor was the tadpole one, {\it i.e.}, 3 instead of 0.5. The cancellation is so large that it leads to the question if we could not altogether forget the chiral effects as well as the nucleonic response.
The answer is definitely no. The sigma propagator is not the whole story and there are important three-body forces which involve a different combination, with less cancellation, as  shown below. In the expression (\ref{MSTAR}) for the nucleon mass evolution we perform 
a field transformation introducing  a new scalar  field, $u=s \,+\, (\kappa_N \,s^2/2g_S)$, such that only a linear term in $u$ enters  this evolution.
 Expressed in term of the $u$ field, the chiral mexican hat potential takes the form~:
\begin{equation}
V^{chiral}=  V={ m_\sigma^2\over 2} \left(s^2+{s^3\over f_\pi} +... 
\right)=
 {m_\sigma^2\over 2} \left(u^2 +{ u^3\over f_{\pi}}(1-2C) + ...\right).
\end{equation}
where we introduce the dimensionless parameter $C=(\kappa_{N}f_{\pi})/(2\,g_S)$. In the formulation with the $u$ field the three-body forces are totally contained in the $u^3$ term of this potential~:
 \begin{equation}
V^{three-body}=  {m_\sigma^2\over 2} \,{\bar u^3\over f_{\pi}}\,(1\,-\,2\,C). 
\label{V3}
\end{equation}
As $\bar u<0$, this force is repulsive for $C>1/2$, which is actually the case in our fit. Without confinement, {\it i.e.,} for $C=0$,  the chiral potential alone  leads to attractive three-body forces. The balance between the effects of the chiral potential and of the nucleonic response is  not the same in the propagation of the scalar field and in the three-body forces. In the first case the amplitude $T^{total}_{\sigma N} = 3\,g_S/f_{\pi}\, +\, \kappa_{N} = (3\,g_S/f_{\pi})\, (1\,-\, 2\,C/3)$, while in the three-body forces  the combination  is $1\,-2\,\,C$. With $C$  of the order one, which is the value in our fit, the cancellation of the tadpole term  is nearly complete in $T^{total}_{\sigma N}$,  while there is an overcompensation  in the three-body potential which becomes repulsive. The existence of repulsive three-body forces in relativistic theories is strongly supported by the nuclear phenomenology  \cite{LKR97,GMST06}. They  play an important role in the saturation.

Going beyond the Hartree approximation where the pion does not contribute we have introduced  \cite{CE07} its contribution which occurs via the Fock term and the correlation one (including $\Delta$ excitations). For consistency we have also introduced the rho meson exchange. With the introduction of the pion (or rho) exchange interaction it is necessary to consider also the short range interaction in the spin-isospin channels in the form of contact interactions governed by the Landau Migdal parameters, $g'_{NN}, g'_{N\Delta}, g'_{\Delta \Delta}$, where the indices refer to the type of hole or particle. 

Our fit which leads to a successfull description of the nuclear binding goes as follows.

{\it a)} For the quantity  $g_S /m_{\sigma}^2$ we take the lattice value: $g_S /m_{\sigma}^2\simeq a_2/f_\pi= 15\, GeV^{-2}$.
This corresponds, to leading order in density, to a mean scalar field of $20\, MeV$ at normal density. We also need separately the scalar coupling constant $ g_S$, for which we keep the value of the linear sigma model, $g_S=M_N/f_{\pi}$~, although confinement may introduce some  deviation.  The corresponding  $\sigma$ mass is then $\simeq 800 \,MeV$.

{\it b}) The quantity $C=(\kappa_{N}\,f_{\pi})/(2\,g_S)$ which determines the nucleonic response is allowed to vary near the lattice value, $C_{lattice}=1.25$. The fit gives a value which is close, $C=1$.

{\it c}) For the vector part of the potential, the omega mass has the experimental value, $m_{\omega}= 783 \,MeV$, while
 the coupling constant $g_{\omega}$ is a free parameter. The fit value, $g_{\omega}\simeq 7$, is close to the vector dominance quark model one, 
 $g_{\omega}\simeq 8$.

{\it d}) We take for the $\pi N$ form factor  a dipole form with $\Lambda=0.98 \,GeV$. It leads to a pionic sigma commutator 
 $\sigma_N^{pion}=21 \,MeV$ which, added to the non pionic value from the lattice $\Sigma_N^{non-pion}=29\, MeV$, gives a total value $\Sigma_N=50 \,MeV$, in agreement with the experimental value.

{\it e}) For the Landau-Migdal parameters of the spin-isospin interaction we use the latest information from spin-isospin physics \cite{ISW06}~:
 $g'_{NN}=0.7, g'_{N\Delta}=0.3, g'_{\Delta \Delta}=0.5$,  producing  a large suppression of the  correlation terms. 
\\

In summary we have found a full consistency between lattice QCD and the binding properties, relying  on the following assumptions.
The chiral invariant nuclear scalar  field  affects the quark condensate. The nucleon mass has a mixed origin, in part from confinement and in part from the quark condensate. With this limited set of assumptions it is possible to fix or constrain the scalar parameters by  the lattice expansion ones, leading to  a successful description of the nuclear binding properties. Confinement is an essential ingredient to reach saturation. It limits the attractive effect of the chiral mexican hat potential. The consistency 
between lattice data and the saturation properties, which is not {\it a priori} acquired, confirms the picture that the scalar potential which binds the nucleus reflects the modification of the QCD vacuum in the nuclear medium. This is true not only at the level of the mean scalar field 
related to the quark condensate evolution, pionic component removed, but also at the level of the three body forces, which are governed by the nucleonic QCD scalar susceptibility.

This consistency between QCD and the nuclear potential is in itself remarkable  but one may wonder wether there is a deeper necessity and significance for the introduction of both the tadpole term and the confinement one in the sigma propagation.  After  preparation of this talk I read an article by Delbourgo and Scadron \cite{DS95} which may contain the answer to this question. In this work they introduce the linear sigma model at the quark level. Their lagrangian has a simple form which does not include the mass terms for
 pion and sigma, nor the mexican hat potential. These are generated dynamically by quark loops. In particular for the sigma the two graphs which build the $\sigma$ mass are the tadpole term involving a quark loop and the creation of a quark and antiquark pair by the $\sigma$ (fig. 1a and 1b). These are constituent quarks whose masses  obey a gap equation. This is the vacuum situation. Consider now how these two contributions are modified in the nuclear medium. The first one is changed by the extra tadpole term from the constituent quarks of the nucleons (fig. 1c). As for the second one, the creation of a quark antiquark pair by the sigma field, through the interaction of the sigma with a constituant quark of the nucleon, as shown in the graph 1d, represents precisely the nucleonic response arising from confinement which comes from Z graphs. These two terms have  opposite signs. This shows that the two contributions that we have considered in the sigma propagation in nuclei go together and are necessary as they represent the medium modifications of the two contributions which build the sigma mass in the vacuum.

\begin{figure}
  \includegraphics[height=.2\textheight]{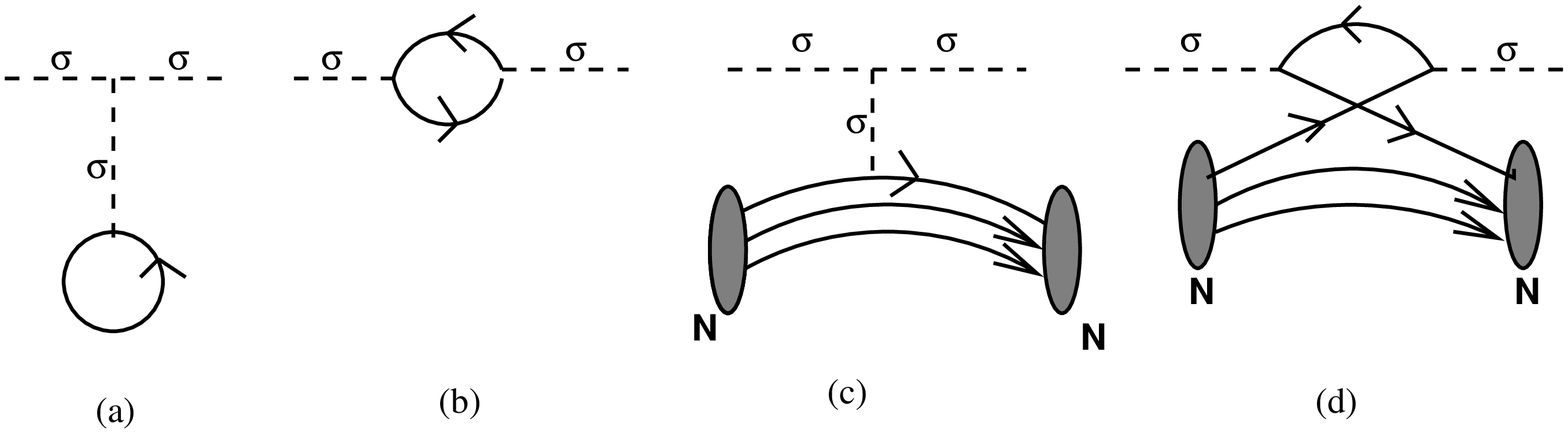}
  \caption{Origin of the sigma mass in the vaccuum from ref. \cite{DS95} (1a and 1b) and its modifications in the nuclear medium (1c and 1d). The quark lines (continuous lines) represent constituent quarks.} 
\end{figure}

\vskip 1 true cm
\noindent
{\it Acknowledgement.} M. E.   thanks the organizers for their invitation to this meeting and for the support.



\begin{thebibliography}{9}
\bibitem{SW86} B.D. Serot, J.D. Walecka, Adv. Nucl. Phys. 16(1986) 1; Int. J. Mod. Phys. E16, 15 (1997).
\bibitem{G88} P.A.M. Guichon, Phys. Lett. B200 (1988) 235.
\bibitem{GSRT96} P.A.M. Guichon, K. Saito, E. Rodionov and A. W. Thomas, 
Nucl. Phys. A601 (1996) 349.
\bibitem{EC07}M. Ericson and G. Chanfray, EPJA 34 (2007) 215.
\bibitem{B96} M. Birse, Phys. rev. C43, (1996) 2048. Acta Phys. Pol. B29,  (1998) 2357.
\bibitem{CEG01} G. Chanfray, M. Ericson, P.A.M. Guichon, Phys. Rev C63 
(2001) 055202.
\bibitem{KM74} A.K. Kerman and L.D. Miller in ``Second High Energy Heavy 
Ion Summer Study'', LBL-3675, 1974. 
\bibitem{CE05} G. Chanfray, M. Ericson, EPJA25 (2005) 151.
\bibitem{CEG03} G. Chanfray, M. Ericson, P.A.M. Guichon, Phys. Rev C68 
(2003) 035209.
\bibitem{TLY04}A. W. Thomas, D. B.  Leinweber and R. D. Young,
Phys. Rev. Lett. 92 (2004) 242002.
\bibitem{TGLY04} A. W. Thomas, P. A. M. Guichon, D. B.  Leinweber and R. 
D. Young, Progr. Theor. Phys. Suppl. 156 (2004) 124; nucl-th/0411014.
\bibitem{ST99}  H. Shen and H. Toki, Phys.Rev. C61 (2000) 045205.
\bibitem{LKR97} G.A. Lalazissis, J. Konig and P. Ring, Phys. Rev. C55 (1997) 540.
\bibitem{GMST06} P.A.M Guichon, H.H. Matatevosyan, N. Sandulescu, A.W. Thomas, Nucl.Phys. A772 (2006) 1.
\bibitem{CE07} G. Chanfray  and  M. Ericson, Phys.Rev. C75 (2007) 015206.
\bibitem{ISW06} M. Ichimura, H. Sakai and T. Wakasa, Prog. Part. Nucl. Phys., 56, 446 (2006).
\bibitem{DS95} R. Delbourgo and M. Scadron, Mod. Phys. Lett. A10(1995) 251.



\end{thebibliography}
\end{document}